\documentclass[12pt,preprint]{aastex}

\def\kms{km~s$^{-1}$}

\def\gs{$\sigma$}

\def\kms{\mbox{km~s$^{-1}$}}
\def\cmc{cm$^{-3}$}

\def\deg{$^{\circ}$}
\def\Msun{\mbox{$M_\odot$}}
\def\Lsun{\mbox{$L_\odot$}}
\def\Rsun{\mbox{$R_\odot$}}
\def\Vlsr{$V_{\rm LSR}$}
\def\Vsys{$V_{\rm sys}$}

\def\amm{NH$_3$}

\def\hcop{\mbox{HCO$^+$}}

\def\HtCOp{H$^{13}$CO$^+$}

\def\dVres{\mbox{$\Delta v_{\rm res}$}}
\def\dVint{\mbox{$\Delta v_{\rm int}$}}

\def\V0{\mbox{$V_{\rm 0}$}}

\def\Vpblue{\mbox{$V_{\rm p,blue}$}}
\def\Vpred{\mbox{$V_{\rm p,red}$}}
\def\Vinf{\mbox{$V_{\rm inf}$}}

\def\dVnth{\mbox{$\Delta v_{\rm nth}$}}

\def\Tr{\mbox{$T_{\rm rot}$}}
\def\Tr21{\mbox{$T_{\rm r,21}$}}
\def\Tmb{\mbox{$T_{\rm mb}$}}

\def\Tmb{\mbox{$T_{\rm mb}$}}

\def\Ciso{$c_{\rm iso}$}

\def\nurest{\mbox{$\nu_{\rm rest}$}}

\def\taut{$\tau_{\rm tot}$}

\def\gf{\mbox{GF\,9-2}}

\def\thetahpbw{$\theta_{\rm HPBW}$}

\def\RadpSimSoltn{$\rho (r)\propto r^{-2}$}

\def\lesssim{\mathrel{\hbox{\rlap{\hbox{\lower4pt\hbox{$\sim$}}}\hbox{$<$}}}}
\def\gtrsim{\mathrel{\hbox{\rlap{\hbox{\lower4pt\hbox{$\sim$}}}\hbox{$>$}}}}

\slugcomment{Accepted by ApJ Letters on 2009 January 6}

\shorttitle{Evidence for Infall in \gf}
\shortauthors{Furuya et al.}

\begin{document}


\title{Spectroscopic Evidence for Gas Infall in \gf}


\author{Ray S. FURUYA\altaffilmark{1}}
\affil{Subaru Telescope, National Astronomical Observatory of Japan}
\email{rsf@subaru.naoj.org}

\author{Yoshimi KITAMURA\altaffilmark{2}}
\affil{Institute of Space and Astronautical Science, Japan Aerospace Exploration Agency}
\email{kitamura@isas.jaxa.jp}

\and

\author{Hiroko SHINNAGA\altaffilmark{3}}
\affil{Caltech Submillimeter Observatory, California Institute of Technology}
\email{shinnaga@submm.caltech.edu}


\altaffiltext{1}{650 North A'ohoku Place, Hilo, HI\,96720}
\altaffiltext{2}{3-1-1 Yoshinodai, Sagamihara, Kanagawa 229-8510, Japan}
\altaffiltext{3}{111 Nowelo Street, Hilo, HI\,96720}


\begin{abstract}
We present spectroscopic evidence for infall motion of gas 
in the natal cloud core harboring an extremely young 
low-mass protostar \gf.
We previously discussed that the ongoing collapse of the \gf\
core has agreement with the Larson-Penston-Hunter (LPH)
theoretical solution for the gravitational collapse of a core
(Furuya et al.; paper I).
To discuss the gas infall on firmer ground, we have
carried out On-The-Fly mapping observations of the \hcop\ (1--0) line 
using the Nobeyama 45\,m telescope equipped with the 
25 Beam Array Receiver System.
Furthermore, we observed the HCN (1--0) line with the 45\,m telescope, 
and the \hcop\ (3--2) line with the
Caltech Submillimeter Observatory 10.4\,m telescope.
The optically thick \hcop\ and HCN lines
show blueskewed profiles whose deepest absorptions are seen 
at the peak velocity of optically thin lines, i.e., 
the systemic velocity of the cloud (paper I), 
indicating the presence of gas infall toward the central protostar.
We compared the observed \hcop\ line profiles with model ones by 
solving the radiative transfer in the core under LTE assumption.
We found that the core gas has a constant 
infall velocity of $\sim$0.5 \kms\ 
in the central region, 
leading to a mass accretion rate of 
2.5$\times 10^{-5}$ \Msun\ yr$^{-1}$.
Consequently, 
we confirm that the gas infall in the \gf\ core is consistent 
with the LPH solution. 
\end{abstract}





\section{Introduction}
\label{s:intro}

Without accurate knowledge of the collapsing process
of an isolated dense cloud core, 
we cannot understand how a low-mass protostar forms
through the accretion process.
One of the major limiting factors is that the initial conditions 
of core collapse are not necessarily revealed
by observations in detail. 
There has been a long standing theoretical debate about
the gravitational collapse process since the early 1970s. 
Two extreme models of runaway collapse and quasi-static, 
inside-out collapse have been developed;
the former is originally proposed by the
Larson (1969) and Penston (1969) and extended by Hunter (1977), 
and the latter by Shu (1977).
In this paper, we refer to the runaway collapse scenario
as the Larson-Penston-Hunter (LPH) solution
by following the nomenclature in McKee \& Ostriker (2007).
The validity of these solutions has been observationally 
studied by e.g., 
Ward-Thompson et al. (1994), and 
Looney, Mundy \& Welch (2003).

In this context, we performed a detailed study of 
the natal cloud core harboring 
an extremely young low-mass protostar \gf\ 
(Furuya, Kitamura \& Shinnaga 2006; hereafter paper I).
The protostar is believed not to have generated an extensive
molecular outflow, yielding a rare opportunity to investigate 
core collapse conditions free from the disturbance by the outflow.
We discussed that the observed velocity field of 
the core is consistent with the initial conditions 
assumed in the LPH solution.
The core shows a radial density profile of 
\RadpSimSoltn\ and the entire core has non-thermal line widths
of \dVnth $\sim$ (2--3)\Ciso, 
suggestive of gas infall motions
(\Ciso\ is isothermal sound velocity).
In order to shed light on the physical properties of the infall motions,
we have carried out molecular line 
observations of the \gf\ core to detect the blueskewed profiles
characterizing infall motion
(e.g., Walker et al. 1986; Zhou et al. 1993; 
Ward-Thompson et al. 1996;
Myers et al. 1996;
Mardones et al. 1997; 
Onishi, Mizuno \& Fukui 1999; Remijan \& Hollis 2006).

\section{Observations}
\label{s:obs}

We have carried out On-The-Fly (OTF) mapping observations
(Sawada et al. 2008) of the \hcop\ $J=$1--0 line 
[rest frequency ($\nu_{\rm rest}$) =\,89188.526\,MHz] 
using the Nobeyama Radio Observatory 
(NRO)\footnote{Nobeyama Radio Observatory 
is a branch of the National Astronomical Observatory of Japan, 
National Institutes of Natural Sciences.} 
45\,m telescope in 2006 May.
We used the 25 Beam Array Receiver System (BEARS), and configured
auto-correlators (ACs) as a backend, 
yielding an effective velocity resolution
(\dVres ) of 0.048 \kms\ in the 8\,MHz bandwidth mode.
At the line frequency, the mean beam size (\thetahpbw ) of the 25 beams was 
18\farcs6, and the mean main-beam efficiency ($\eta_{\rm mb}$) was 0.51.
All the spectra were calibrated by the standard chopper wheel method, 
and were converted into main-beam brightness temperature
(\Tmb ) by dividing by $\eta_{\rm mb}$.
We adopted the correction factors provided by the observatory for
gain differences between the 25 beams.
The uncertainty in our intensity calibration is estimated to be $\sim$\,15\%. 
The OTF mapping was carried out over an area with a size of $\sim$320\arcsec,
employing position switching method.
The telescope pointing was checked every 1.2 hrs, and was
found to be accurate within 3\arcsec.
The data reduction was done using the NOSTAR package.
By using a spheroidal function as a gridding convolution function,
we produced a data cube with a spatial resolution of 
20\farcs6 and a pixel size of 5\arcsec.
Furthermore, in 2008 June, we have performed a single-point 
integration of the HCN (1-0) line toward the core center 
with the 45\,m telescope. 
The beam size of the S80 receiver at the line frequency was 
18\farcs2 in HPBW, 
and $\eta_{\rm mb}$ was 0.44.
We used acousto-optical spectrometers (AOSs), AOS-H, 
providing \dVres\ of 0.13 \kms.\par

In 2006 July, 2007 August, and 2008 June,
we have used the Caltech Submillimeter
Observatory (CSO)\footnote{Caltech Submillimeter Observatory is operated by
the California Institute of Technology under the grant from
the US National Science Foundation (AST 05-40882).} 
10.4\,m telescope to observe the \hcop\ $J=$3--2 transition
($\nu_{\rm rest}=$\,267557.633\,MHz).
We assumed \thetahpbw $=$ 26\arcsec\ at 267\,GHz 
and estimated $\eta_{\rm mb}$ of $0.70\pm 0.15$ 
from our measurements of Jupiter.
We configured the 50\,MHz band width AOS,
providing \dVres\ of 0.055 \kms.
The \hcop\ line was observed toward 3$\times$3 positions with a grid
spacing of 30\arcsec\ centered on the \gf\ core center.
The pointing accuracy was better than 3\arcsec, 
and the overall uncertainty in flux calibration was 30\,\%.

\section{Results}
\label{s:results}

\begin{figure}
\begin{center}
\includegraphics[angle=0,scale=.65]{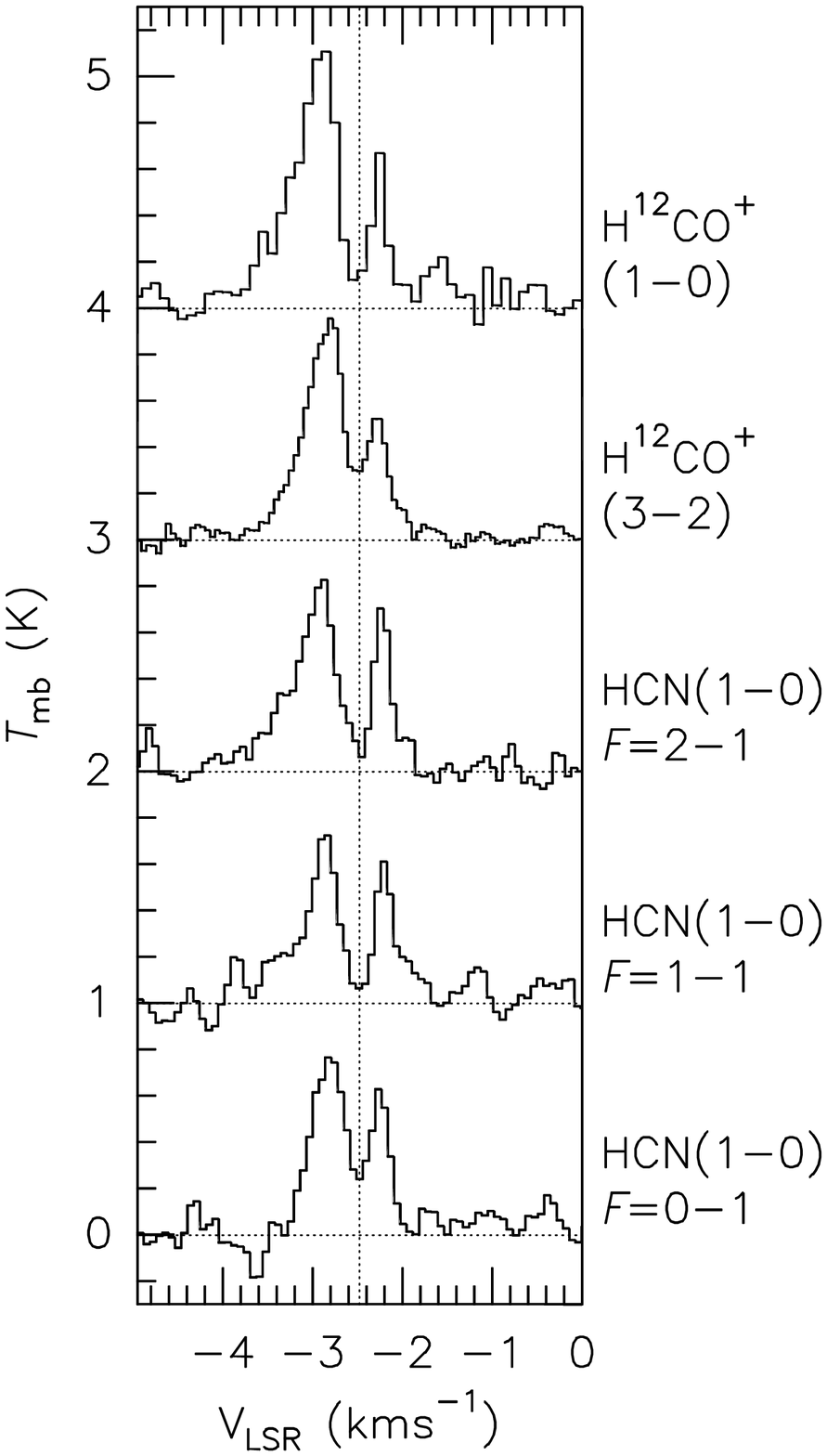}
\caption{Single-dish spectra of the molecular line emission toward the
\gf\ core center shown in the main-beam brightness temperature (\Tmb) scale. 
The \hcop\ (1--0) spectrum is smoothed 
so as to have the effective 
resolution of 26\arcsec\ in FWHM,
i.e., the CSO beam size for the \hcop\ (3--2) line.
The RMS noise levels are
73\,mK, 61\,mK, and 88\,mK for the
\hcop\ (1--0), \hcop\ (3--2), and the HCN lines, respectively.
To convert the rest-frequency (\nurest\ ) 
of the HCN hyperfine emission 
to the LSR-velocity (\Vlsr ),
we adopted the systemic velocity (\Vsys ) of $\,-2.48$ \kms\ (paper I),
and \nurest\ $=$ 
88630.41406\,MHz for $F=$\,1--1,
88631.84375\,MHz for $F=$\,2--1, and
88633.93750\,MHz for $F=$\,0--1.
The vertical dashed line shows
the systemic velocity of the cloud, \Vsys\ $=\,-2.48$ \kms\ (paper I).
\label{fig:sp}}
\end{center}
\end{figure}

Figure \ref{fig:sp} shows molecular line spectra toward the \gf\ core 
center in unit of \Tmb.
All the spectra, except the \hcop\ $J=1-0$, 
are the single-dish spectra taken with position switching. 
The \hcop\ (1--0) spectrum was produced 
from the data cube ($\S\ref{s:obs}$) 
by convolving with a Gaussian profile in order
to have an effective spatial resolution of 26\arcsec\ in FWHM, 
which is equal to the CSO beam size for the \hcop\ (3--2) line.
Clearly, all the spectra show blueskewed profiles.
The LSR-velocities of their deepest absorption agree with
the systemic velocity (\Vsys) of the cloud, \Vlsr $= -2.48$ \kms,
which was estimated from the optically thin
\HtCOp\ (1--0) and CCS $4_3-3_2$ lines (paper I)
as well as \amm\ inversion lines at 23\,GHz 
(Furuya, Kitamura \& Shinnaga 2008).\par

The velocities of the blueshifted peaks of the \hcop\ (1--0) 
and (3--2) lines, $V_{\rm p,blue}$ are estimated to be 
$-2.96\pm 0.02$ \kms\ and $-2.86\pm 0.01$ \kms, respectively, 
from Gaussian fitting. 
Similarly, we obtained $V_{\rm p,red} = -2.29\pm 0.01$ \kms\ 
and $-2.25\pm 0.05$ \kms\ 
for the redshifted peaks of the (1--0) and (3--2) transitions, 
respectively.
The obtained \Vpred\ for both the transitions agree with 
each other, while the \Vpblue\ values differ by 0.1 \kms\ for the
two transitions, suggesting that the velocity difference should
be attributed to the intrinsic property of the gas rather than 
uncertainties in \nurest.\par

The three hyperfine (HF) components of the HCN (1--0) emission, 
showing blueskewed profiles, 
have comparable peak intensities, suggestive of large optical depths.
We have performed HFS analysis of the blue- and redshifted 
components separately assuming that infall motions traced by the 
blue- and redshifted emission are independent, 
and that each emission can be approximated by a Gaussian profile.
Our analysis gave the representative LSR-velocity 
($V_0$) of $V_{0,\rm blue} = -2.90\pm 0.01$ \kms, 
intrinsic velocity width (\dVint ) of $0.34\pm 0.01$ \kms, and
total optical depth (\taut ) 
of $15\pm 2$ for the blueshifted components.
Similarly, we obtained
$V_{0,\rm red} = -2.21\pm 0.03$ \kms, 
\dVint\ $= 0.17\pm 0.02$ \kms,
and \taut $= 12\pm 1.6$ for the redshifted components.
These \taut\ values lead to optical depths ($\tau$) of 1.6 and 1.3 
for the blue- and redshifted $F=0-1$ components, 
8.1 and 6.7 for the $F=2-1$ components, and 
4.9 and 4.0 for the $F=1-1$ components, respectively, 
with uncertainty of 20\%.
The estimated $\tau$ for the blue- and redshifted gas 
of each HF component are comparable to each other,
as expected from their intensity ratios,
and seem to be consistent with 
the fact that the absorption dips in the $F=2-1$ and $F=\,1-1$ 
transitions are deeper than that in the $0-1$ transition.\par

Figure \ref{fig:totmap} compares total integrated intensity maps 
of the \HtCOp\ (1--0) (paper I) and \hcop\ (1--0) emission.
Both lines show a similar spatial structure, but
the \hcop\ emission is more extended than the \HtCOp.
Figure \ref{fig:pmap} represents the \hcop\ (3--2) and (1--0) mosaic spectra 
taken with 30\arcsec\ spacing. 
Here, the (1--0) spectra were produced
from the smoothed cube data 
in the same manner as for Figure \ref{fig:sp}.
Both the transitions show the blueskewed profiles 
not only toward the core center, 
but also at some surrounding positions of,
e.g.,  
($\Delta\alpha, \Delta\delta$) $=$ 
($0\arcsec, ~\pm 30\arcsec$) and
($+30\arcsec, ~\pm 0\arcsec$).
Due to the limited S/N, 
it is difficult to assess
whether or not the outer spectra 
(e.g., in the row of $\Delta\delta = +60\arcsec$
and the column of $\Delta\alpha = +60\arcsec$)
have blueskewed profiles.
The \hcop\ (1--0) emission show strong blueshifted emission 
toward the southwest, 
leaving a possibility that the emission represents 
another blueshifted component of the gas.
In fact, the possibility is supported by Position-Velocity 
(PV) diagrams of the \hcop\ (1--0) and 
\HtCOp\ (1--0) emission shown in Figure \ref{fig:pv}. 
Clearly, there exists an isolated blueshifted component of 
the \hcop\ emission peaked at the offset position of 
$\sim -120\arcsec$ (Figure \ref{fig:pv}a). 
It should be noted that the velocity structures in the 
central region of $r \lesssim 30\arcsec$
are fairly consistent with the blueskewed profiles
expected for infall motion.

\section{Discussion}
\label{s:discussion}

\begin{figure}
\begin{center}
\includegraphics[angle=-90,scale=.7]{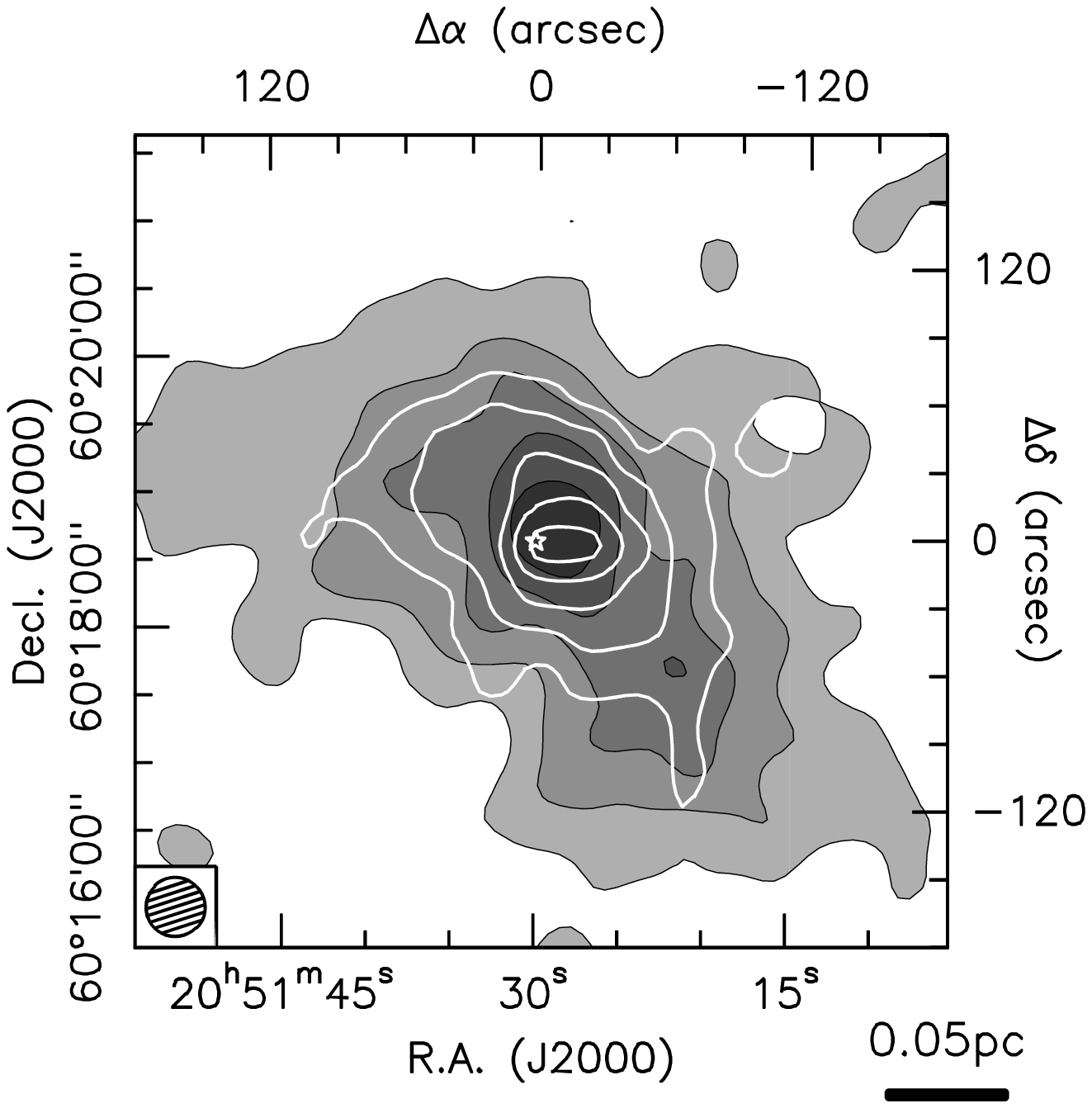}
\caption{Overlay of the total integrated intensity maps of the
\HtCOp\ (1--0) emission (white contour; paper I) on that of the 
\hcop\ (1--0) emission
(grey-scale with thin contour; this work) emission.
The contour levels for the \hcop\ emission
have the 3\gs\ intervals starting from the 3\gs\ level
where \gs\ $=$ 0.59 K \kms.
The \hcop\ emission is integrated over an LSR-velocity range
between $-3.60$ \kms\ and $-1.95$ \kms.
The central star marks the peak position of the 3\,mm continuum
emission (paper I). The hatched circle at the bottom left
corner indicates the effective beam size of the \hcop\ emission,
which is set to be equal to the CSO beam size for \hcop\ (3--2)
(\thetahpbw\ $=$\,26\arcsec; see $\S\ref{s:obs}$).
}
\label{fig:totmap}
\end{center}
\end{figure}

\begin{figure}
\begin{center}
\includegraphics[angle=-90,scale=.80]{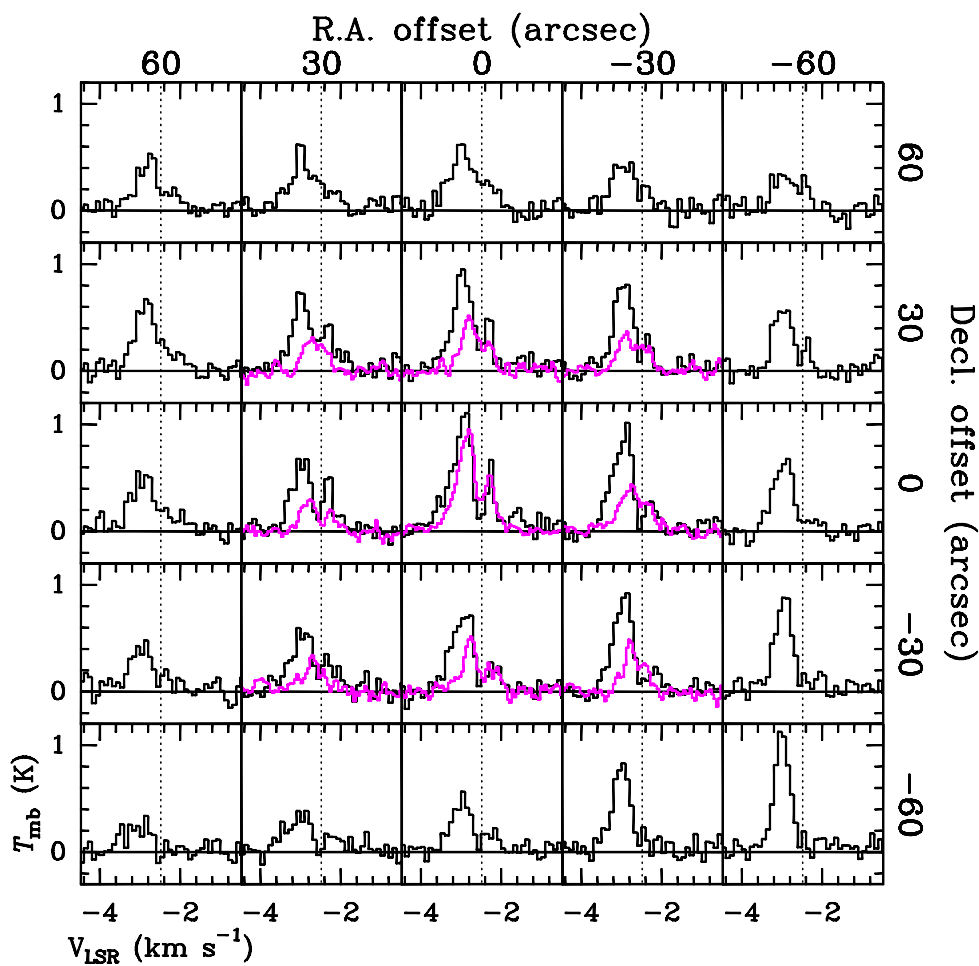}
\caption{Mosaic spectra of the \hcop\ (1--0) (black histogram) and
\hcop\ (3--2) (purple histogram) lines in main-beam temperature
(\Tmb ) scale with 30\arcsec\ spacing centered on 
on the 30\arcsec\ spacing grid centered at
the \gf\ core center.
The vertical dashed line in each panel indicates \Vsys.
\label{fig:pmap}}
\end{center}
\end{figure}

\begin{figure}
\begin{center}
\includegraphics[angle=-90,scale=.45]{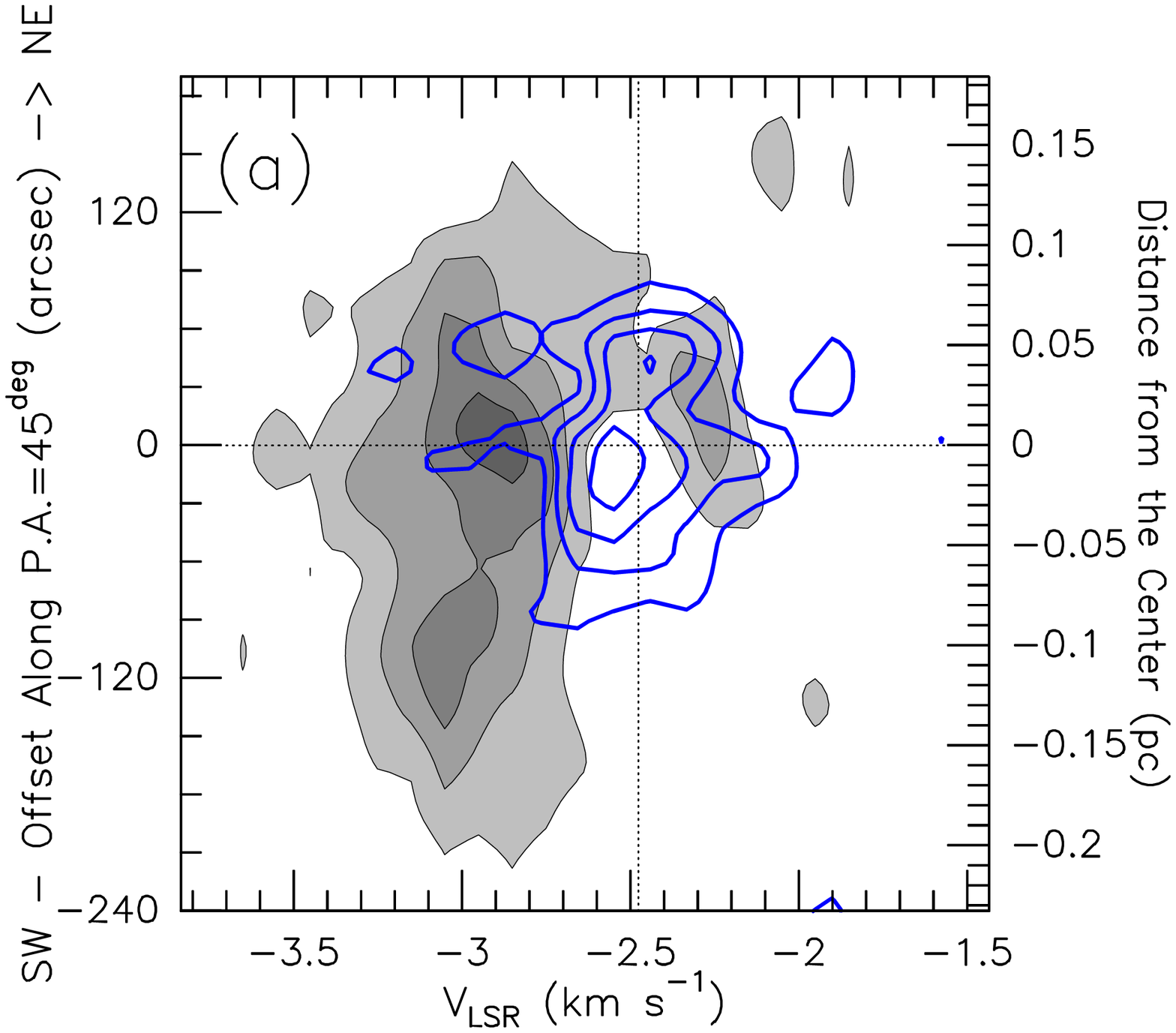} \\
\includegraphics[angle=-90,scale=.45]{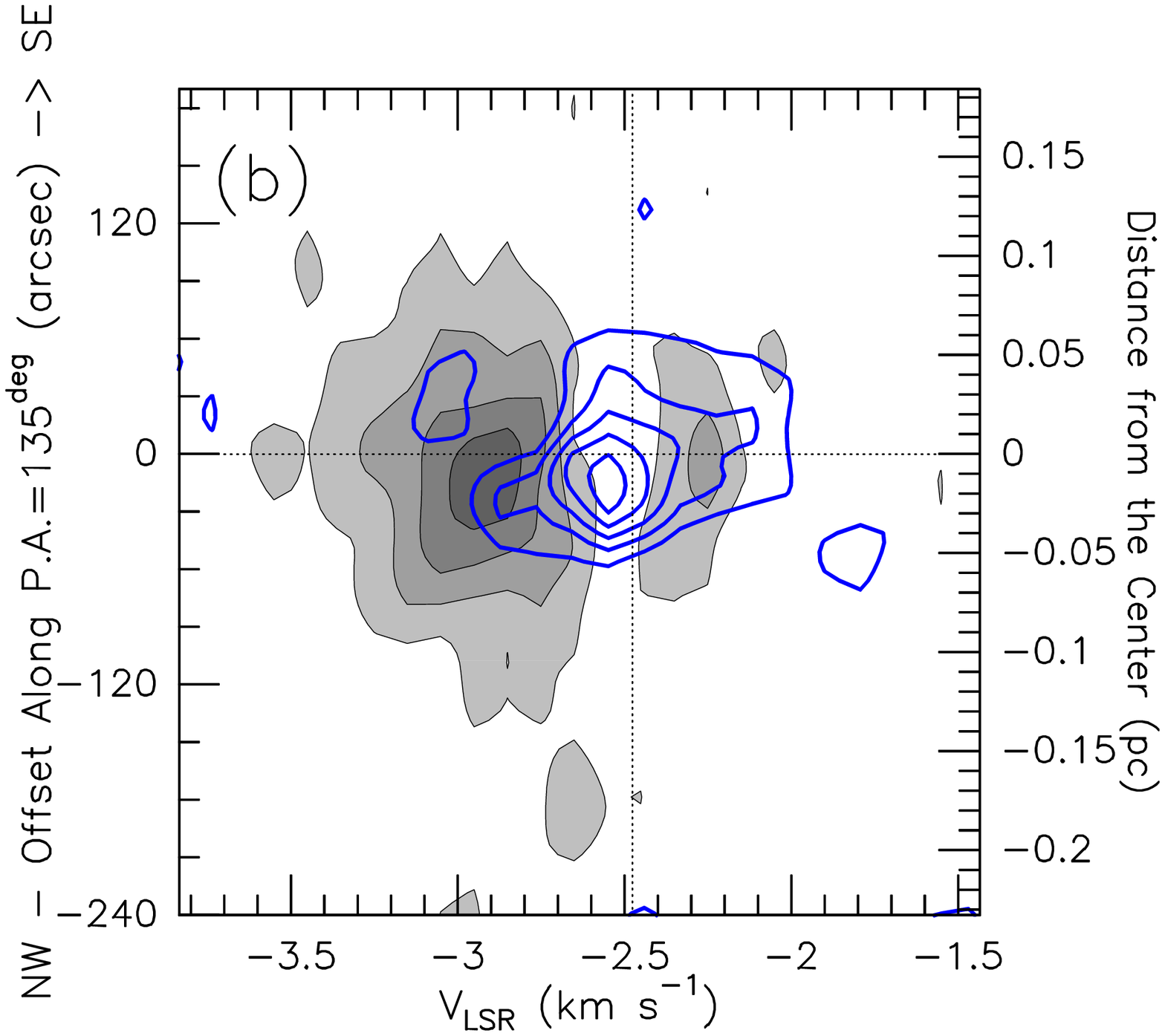}
\caption{
Position-Velocity (PV) diagrams of the 
\hcop\ (1--0) (grey-scale with thin contour; this work) and
\HtCOp\ (1--0) (blue; paper I) emission along
(a) the major (P.A.$=+45$\deg) and
(b) minor (P.A.$=135$\deg) axes passing the core center.
Both the line data are smoothed with the 26\arcsec\ beam 
the same as for Figure \ref{fig:totmap}.
The contour levels are the 3$\sigma$ interval starting
from the 3$\sigma$ level of each emission where
$\sigma =$ 131 and 57 mK per velocity channel
in \Tmb\ for the \hcop\ and \HtCOp\ emission, respectively.
The vertical dashed line in each panel indicates \Vsys.
\label{fig:pv}}
\end{center}
\end{figure}

Although the nature of the blueshifted \hcop\ emission to the
southwest is not clear, 
our observations show that the gas motion in the central
region ($r \lesssim 30\arcsec$) is governed by the infall.
Therefore, we focus on the discussion of the spectrum toward 
the core center.
To give more quantitative constraints on the gas infall,
we have developed a radiative transfer code to model the H$^{12}$CO$^+$
and \HtCOp\ spectra
for a spherical core dominated by infall motion.
The code was originally written for the CO emission 
from a protoplanetary disk (Omodaka, Kitamura \& Kawazoe 1992).
In this model, we assumed that local thermodynamic equilibrium (LTE) 
holds for the line excitation.
To calculate the spectra, 
we assumed that the gas is infalling with a constant velocity 
($v=$\,\Vinf) for $r\leq$ 30\arcsec, whereas a static state ($v=0$) 
for 30\arcsec $<r\leq$ 100\arcsec\ from 
Figures \ref{fig:totmap}--\ref{fig:pv}.
We adopted the mass density radial profile reported in paper I: 
$\rho(r) = 8.4\times 10^{23}(r/{200\,\rm AU})^{-2}$ \cmc.
We used a radial temperature profile of 
$T(r) = 1000\,(r/{\rm 1\,AU})^{-p}$\,K with $p=1/2$,
on the basis of the temperature profile at $2\times 10^4$ yr 
after the formation of a protostar in Figure 2 by \citet{ms00b}.
We took a constant fractional abundance for 
\HtCOp\ of $(8.5\pm 4.7)\times 10^{-12}$ (paper I)
all over the core, although this simplification may not be valid in a 
dynamically collapsing core (e.g., Aikawa et al. 2001).
For a comparison with the observed spectra, 
we smoothed the calculated ones with a Gaussian function with
\thetahpbw\ $=$ 26\arcsec\ and 
converted the flux density into the \Tmb\ 
(see Omodaka et al. 1992).\par

\begin{figure}
\begin{center}
\includegraphics[angle=-90,scale=.58]{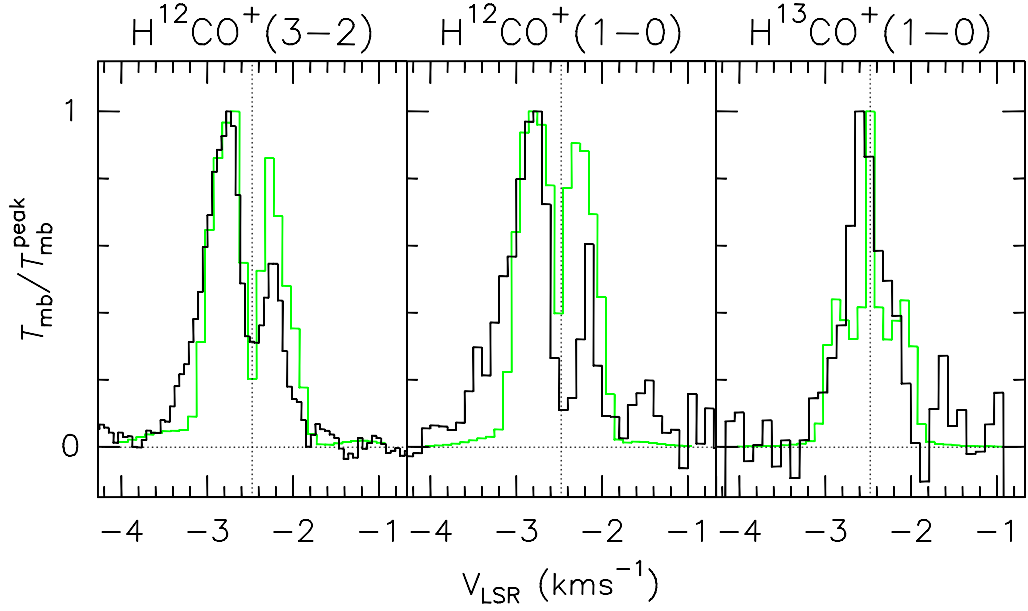}
\caption{
Model (green) and observed (black) spectra of the 
H$^{12}$CO$^+$ (3--2), H$^{12}$CO$^+$ (1--0), and \HtCOp\ (1--0) emission
toward the center of the core.
The model spectra are calculated with the most plausible infall velocity 
(\Vinf ) of 0.5 \kms\ (see text).
To compare with the observed profiles,
the model spectra are convolved by a Gaussian
function with \thetahpbw\ $=$ 26\arcsec\ 
(see $\S\ref{s:discussion}$).
\label{fig:mdl}}
\end{center}
\end{figure}

Figure \ref{fig:mdl} compares calculated and observed
spectral profiles whose intensity scales are normalized.
Here we discuss only the velocity differences between 
the blue- and redshifted peaks because our calculations adopted the 
LTE approximation, and because rigorous calculations \citep{ms00b}
show that the intensities of the model spectra significantly vary 
with the evolutionary stage of the protostar.
We searched for a plausible value of \Vinf\ to reproduce 
the observed line profiles, and found that 
a velocity of 0.5 \kms\ gives the most plausible H$^{12}$CO$^+$ line profiles.
In fact, the calculated H$^{12}$CO$^+$ spectra have a reasonable consistency 
with the observed ones within the errors (Figure \ref{fig:mdl}).
We confirmed that the two cases of \Vinf\ $=$ 0.3 and 0.7 \kms\ 
cannot reproduce the observed velocity differences 
between the blue- and redshifted peaks ($\S\ref{s:results}$).
Notice that \Vinf\ $=$ 0.5 \kms\ agrees
with an estimate from \Vinf\ 
$\approx\frac{1}{2}|V_{\rm blue} - V_{\rm red}|/\cos(45^{\circ}) =$
\,0.45 \kms\ in $\S\ref{s:results}$.
Although the model \HtCOp\ line profile has the blue and red shoulders 
due to the central infalling region, 
the overall line profile seems agree with the observed one 
that has a single peak at \Vsys\ owing to the outer static gas.
As for $p$, we also tried the two cases of $p =$ 1/3 and 1, 
but found that both the $p$ values cannot reproduce 
the observed absorption features around \Vsys.
Here, the $p$ value of 1/3 is expected in the case that 
the dust opacity coefficient has the frequency 
dependence of $\propto \nu^2$ \citep{doty94}.
We addressed such a case because the overall properties
of the dust in the \gf\ core, 
which is at a very early evolutionary stage, 
may be similar to those for the interstellar medium.
In addition, the deep absorption dips around \Vsys\ require 
the presence of a cold static envelope surrounding 
the contracting region.\par

It is known that a density profile of $\rho(r) \propto r^{-\frac{3}{2}}$ 
is expected for a central free-fall region onto the forming protostar.
Our previous observations revealed that the profile of
$\rho(r) \propto r^{-2}$ continues down to the central 
$r\sim$\,600\,AU region, although
it was impossible to know whether or not 
$\rho(r) \propto r^{-\frac{3}{2}}$ exists at $r\lesssim 600$\,AU owing to 
our insufficient angular resolution.
Given $p =1/2$ and \Vinf\ $=$ 0.5 \kms, 
we further calculated the \hcop\ spectra for the case that 
the gas is freely falling onto the central protostar 
of 0.1 \Msun\ (paper I) at $r<$ 600 AU.
However, the resultant line profiles were not 
significantly changed compared with the above results.\par

Although the LPH solution well describes 
the on-going collapse in the \gf\ core (paper I), 
we examined a case for the Shu's solution.
We assumed that the gas is freely falling toward
a 0.1 \Msun\ protostar at $r\lesssim$\,600 AU and is
static at $r\gtrsim 600$\,AU 
on the basis of the observed density profile (paper I).
The calculated \hcop\ spectra for the two transitions do not show
blueskewed profiles, but double-peaked spectra
having an absorption dip at the \Vsys.
This would be a natural consequence by giving 
the compact free-fall region embedded in the large static core.\par

Assuming that the infall motion is radial,
we can estimate the mass accretion rate 
$\dot{M}_{\rm acc}$ by $4\pi r^2\rho V_{\rm inf}$.
We obtain $\dot{M}_{\rm acc}$ of 2.5$\times 10^{-5}$ \Msun\ yr$^{-1}$
with uncertainty of a factor $\sim 2$.
Since theoretical models of protostar formation
(e.g., Stahler, Shu \& Tamm 1980;
Masunaga \& Inutsuka 2000a) predict that 
the protostar radius, $R_\ast$,
should be $\sim 5$ \Rsun\ for a low-mass protostar,
we can calculate an accretion luminosity ($L_{\rm acc}$) for
the central protostar(s) of
$L_{\rm acc}= 16 \Lsun\ G
(M_{\ast}/0.1 \Msun)
[\dot{M}_{\rm acc}/(2.5\times 10^{-5}\Msun\ yr^{-1})]
(R_{\ast}/5\Rsun)^{-1}$.
This seems reasonable compared with 
the luminosity expected for a protostar at a very
early evolutionary stage 
(see e.g., Figure 7 in Masunaga \& Inutsuka 2000a).
Consequently, our study confirms that the gas in the \gf\ 
core is infalling onto the protostar(s) 
in the central region, 
and that the infalling velocity is consistent 
with the prediction from the LPH solution.
Furthermore, we plan to perform deeper integration of 
these \hcop\ lines to examine the presence of the infalling 
gas in the outer region of the core.

\acknowledgments

R. S. F. gratefully acknowledges Thomas A. Bell, Thomas G. Phillips,
and Ruisheng Peng for their generous help during the CSO observations.
The authors sincerely thank Thomas A. Bell for a critical reading of the 
manuscript at the final stage of preparation. 
This work is partially supported by Grant-in-Aids 
from the Ministry of Education, Culture, Sports, 
Science and Technology of Japan
(No.\,19204020 and No.\,20740113).



{\it Facilities:} \facility{Nobeyama 45\,m telescope}, 
\facility{Caltech Submillimeter Observatory 10.4\,m telescope}

\clearpage



\end{document}